# A Control-Driven Framework for Secure SaaS Onboarding in Regulated Enterprises

Naga Sundeep Krishna Thota[1] and Rithika Dulam[2]
[1]*Independent Researcher, IEEE Member, Media, Pennsylvania, USA — ORCID: 0009-0005-7665-6326*
[2]*Research Scientist, Virginia Tech, Frederick, Maryland, USA — ORCID: 0009-0006-7788-0765*

***Abstract*—As enterprises increasingly adopt Software-as-a-Service (SaaS) platforms for mission-critical functions, onboarding these services has emerged as a complex governance challenge. In regulated environments, SaaS onboarding must address multiple interdependent control domains, including Third-Party Risk Management (TPRM), cybersecurity assessment, Identity and Access Management (IAM), and disaster recovery (DR). These domains are often executed in isolation, resulting in delayed go-lives, duplicated assessments, unclear ownership, and residual operational risk. This paper proposes a control-driven, end-to-end SaaS onboarding framework that integrates TPRM, cybersecurity, IAM, and DR into a unified lifecycle model spanning intake and risk scoping, architecture validation, identity design, resilience assessment, and post-production governance. Key contributions include: (1) a structured onboarding lifecycle emphasizing sequencing and dependency management across control domains; (2) a cross-domain control mapping that highlights failure modes caused by siloed reviews; and (3) practical design patterns for secure connectivity, federated identity, least-privilege access, and shared-responsibility disaster recovery. Unlike prior frameworks that treat these domains independently, this work introduces a formally gate-sequenced, cross-domain lifecycle, the first integrated model that encodes mandatory dependency ordering across all four control domains with traceable evidence artifacts at each stage, directly addressing structural gap responsible for enterprise-owned SaaS failures such as the 2024 Ticketmaster-Snowflake breach.**

## I. INTRODUCTION

Over the past decade, organizations across regulated industries (financial services, healthcare, insurance, energy, and the public sector) have rapidly adopted SaaS platforms for mission-critical functions including human resources, enterprise finance, collaboration, CRM, and data analytics [1]. This shift delivers real operational benefits: faster deployment, reduced infrastructure burden, continuous vendor-managed updates, and elastic scalability. However, it introduces a fundamental rebalancing of control: unlike traditional on-premises deployments where the enterprise owns the full stack, the SaaS model delegates hosting, operations, and application security to the vendor [1]. The enterprise becomes a tenant consumer, unable to apply perimeter-based controls, audit underlying infrastructure, or do more than rely on contractual assurances and third-party attestations to validate the security posture of services upon which critical operations depend.

This structural shift exposes a new class of enterprise risk that existing IT governance frameworks were not designed to manage holistically. The shared responsibility boundary (what the enterprise controls versus what the vendor controls) is frequently ambiguous, inconsistently documented, and subject to misinterpretation by both parties. As SaaS portfolios grow in breadth and criticality, undermanaged adoption becomes a systemic vulnerability in the enterprise security posture.

In current practice, SaaS onboarding is fragmented across siloed functions: Procurement manages commercial engagement; TPRM conducts vendor-level due diligence; Cybersecurity performs security review; IAM handles authentication integration; and BCP/DR teams evaluate operational resilience, each applying domain-specific controls in isolation, with no shared sequencing model, no dependency map, and no unified accountability structure. This fragmentation produces five recurring challenges that the proposed framework directly addresses: (C1) siloed domain reviews with no shared sequencing model, producing duplicated vendor assessments and contradictory risk decisions across the same engagement; (C2) ambiguous shared-responsibility boundaries, where no formal mechanism exists to classify controls as vendor-owned, enterprise-owned, or jointly-owned, creating persistent control coverage gaps; (C3) IAM integration deferred to post-production, leaving native vendor accounts active and ungoverned beyond go-live; (C4) DR planning treated as a documentation exercise rather than operational validation, with vendor-stated RTO/RPO commitments accepted without enterprise-side rehearsal; and (C5) TPRM vendor approval conflated with production readiness, bypassing post-contract tenant configuration review as a distinct control gate[2]. A foundational principle of enterprise risk governance applies throughout: organizations can outsource the service, but never the risk [2], meaning the enterprise remains fully accountable for the regulatory, financial, and reputational consequences of any SaaS security incident, regardless of vendor contractual commitments.

This paper proposes a unified, control-driven SaaS onboarding framework that sequences TPRM, cybersecurity, IAM, and DR activities across a defined lifecycle, directly addressing challenges C1 through C5. The framework is platform-agnostic, applicable to productivity suites, analytics platforms, cloud data warehousing, collaboration tools, and CRM systems, and is designed to: (1)

reduce onboarding cycle time by eliminating redundant assessments and establishing clear dependency sequencing; (2) eliminate duplicated vendor assessments through cross-domain coordination; (3) enforce clear control ownership across all onboarding phases; and (4) support post-production governance through continuous monitoring and periodic re-assessment.

The remainder of this paper is organized as follows. Section II provides background and related work. Section III analyzes the SaaS threat landscape and presents the 2024 Ticketmaster–Snowflake breach as a case study. Section IV presents the six-stage SaaS onboarding lifecycle framework. Section V provides a cross-domain control mapping and sequencing dependency analysis. Section VI presents six reusable design patterns. Section VII provides a reference onboarding checklist. Section VIII discusses applicability, limitations, and lessons learnt. Section IX concludes the paper with future work.

## II. BACKGROUND AND RELATED WORK

### *A. SaaS in Enterprise*

NIST SP 800-145 defines SaaS as a cloud service model in which provider applications run on cloud infrastructure and are accessible via web browser; the consumer does not manage the underlying infrastructure, including network, servers, operating systems, or storage [1]. Unlike Infrastructure as a Service (IaaS) or Platform as a Service (PaaS), where the enterprise retains varying degrees of stack ownership, SaaS limits enterprise control to introduced data and vendor-exposed configuration settings, meaning the enterprise cannot directly inspect platform security, apply endpoint controls to the application runtime, or conduct independent penetration testing without vendor authorization [3]. Key risk challenges include data residency and sovereignty concerns, multi-tenancy risks from logical tenant isolation, vendor concentration risk, and opaque security configurations misaligned with enterprise standards [4], all of which are amplified in regulated industries requiring demonstrable control over data regardless of delivery model.

### *B. Third-Party Risk Management in Regulated Industries*

TPRM is the structured process of identifying, assessing, monitoring, and mitigating external party risk [5]. In regulated industries, TPRM has evolved from an advisory governance function to a formal regulatory requirement. In financial services, TPRM requirements are codified in guidance from the Federal Deposit Insurance Corporation (FDIC), Federal Financial Institutions Examination Council (FFIEC), Federal Reserve Board, Office of the Comptroller of the Currency (OCC), and UK Prudential Regulation Authority [2], [5]. The OCC Bulletin 2023-17, interagency guidance, explicitly requires due diligence proportionate to the risk and complexity of the third-party relationship [5]. Common assurance mechanisms include SOC 2 Type II, ISO/IEC 27001:2022, and ISO 22301:2019 [6], [7], [8]. A critical limitation is that TPRM evaluates vendor-level security programs, not enterprise-specific tenant configurations; a vendor may hold exemplary attestations while a specific tenant operates with critically weak settings [9]. This gap between vendor-level assurance and tenant-level security is a foundational premise of the proposed framework.

### *C. Cybersecurity Frameworks, Identity Governance and Disaster Recovery*

Relevant cybersecurity frameworks, including NIST CSF 2.0 (Govern, Identify, Protect, Detect, Respond, Recover) [10], ISO/IEC 27001:2022 with Annex A supplier and cloud controls [6], and CIS Controls v8 [11], address general security governance but do not model the distinction between pre-contract vendor assessment (whether the enterprise trusts the vendor) and post-contract tenant configuration review (whether the enterprise has configured its use of the vendor platform securely) as a formal sequencing dependency [12]. SaaS IAM relies on federated protocols (SAML 2.0, OIDC/OAuth 2.0) to position the enterprise Identity Provider (IdP) as the authoritative authentication source [13], with Single Sign-On (SSO) centralizing authentication control and Multi-Factor Authentication (MFA) reducing credential compromise risk [14]. Role-Based Access Control (RBAC) enforces least-privilege at the role level, while Privileged Access Management (PAM) governs privileged accounts through session recording, just-in-time provisioning, and password vaulting. Persistent SaaS-specific IAM risks include native vendor accounts surviving federated account termination, orphaned accounts from incomplete offboarding, and over-privileged service accounts that escape routine recertification cycles. In SaaS, the shared responsibility model splits DR obligations: the vendor owns infrastructure-level recovery, while the enterprise owns tenant-level recovery covering identity infrastructure, integrations, business processes, and user communication, governed by ISO 22301:2019 [8], [15]. A recurring failure is accepting vendor-stated Recovery Time Objectives (RTO) and Recovery Point Objectives (RPO) commitments without validating that enterprise-side dependencies recover within the same windows, a gap consistently exposed through targeted tabletop exercises.

### *D. Gaps in Prior Work and Framework Contributions*

Prior literature has addressed specific dimensions of SaaS security governance in isolation. Subramanian and Jeyaraj [16] identify security challenges in cloud computing but do not model cross-domain sequencing dependencies or define pre/post-contract control separation. Almorsy et al. [17] examine collaboration-based cloud security management and found that effective cloud security depends on coordinated responsibilities and collaboration among cloud service providers and enterprises, but propose no remediation lifecycle or IAM integration model. Rose et al. [18] propose zero-trust architecture emphasizing security with continuous authentication, authorization, and validation of each access request, but does not address TPRM or DR integration within an onboarding lifecycle. Hashizume et al. [19] identify key security vulnerabilities in cloud computing, including access control and identity management weaknesses, to highlight major areas of security risk, but provide no structured remediation sequence or formal gate dependency model. CSA [21] identifies and categorizes major cloud security threats to provide organizations with a comprehensive view of emerging and persistent cloud risks, but does not prescribe an integrated, gated onboarding framework applicable to regulated enterprises. (refer Table I). A review of the existing literature indicates that no prior approach: (a) defines formal gate dependencies between TPRM and cybersecurity review; (b) maintains an explicit pre/post-contract control separation; (c) integrates shared-responsibility DR planning within a unified SaaS onboarding lifecycle; or (d) maps all four control domains (TPRM, cybersecurity, IAM, DR) into a single formally sequenced model. The proposed framework addresses all five challenges simultaneously. Table I provides a structured comparison.

TABLE I.

Prior Works vs. Proposed Framework for C1-C5 Challenge Coverage (✓ = fully addressed; ~ = partially addressed; ✗ = not addressed)

| Challenge | Subramanian & Jeyaraj [16] | Almorsy et al. [17] | Rose et al. / NIST ZTA [18] | Hashizume et al. [19] | CSA Top Threats [21] | Proposed Framework (This Work) |
|---|---|---|---|---|---|---|
| C1: Siloed domain reviews / no shared sequencing model | ✗ | ✗ | ~ | ✗ | ~ | ✓ |
| C2: Ambiguous shared-responsibility boundaries | ~ | ~ | ~ | ~ | ~ | ✓ |
| C3: IAM deferred to post-production; native accounts ungoverned | ✗ | ✗ | ~ | ~ | ~ | ✓ |
| C4: DR treated as documentation; RTO/RPO not validated | ✗ | ✗ | ✗ | ✗ | ✗ | ✓ |
| C5: No formal pre/post-contract control separation | ✗ | ✗ | ✗ | ✗ | ✗ | ✓ |

## III. THREAT LANDSCAPE AND CASE STUDY

### *A. The Evolving SaaS Threat Landscape*

SaaS platforms have become high-value targets for threat actors for structurally predictable reasons: they aggregate substantial volumes of sensitive enterprise and customer data in centrally accessible, internet-reachable endpoints; they rely extensively on customer-managed authentication and configuration controls; and they create privileged API access points that, when compromised, can enable large-scale automated data exfiltration [21].

The Cloud Security Alliance Top Threats to Cloud Computing 2024 identifies Advanced Persistent Threats (APTs) and Insufficient Identity, Credentials, Key, and Access Management as leading threat categories [21]. Key attack vectors targeting SaaS platforms include: credential theft through infostealer malware; identity federation misconfigurations creating authentication bypasses; insecure or unrotated API tokens; misconfigured tenant settings exposing data beyond intended access boundaries; inadequate MFA enforcement enabling credential stuffing; and supply chain attacks propagated through SaaS-to-SaaS integration chains [22]. The enterprise-facing implication is significant: the most consequential SaaS security failures are not vendor platform vulnerabilities; they are customer-owned control failures operating within a correctly functioning vendor platform.

### *B. Case Study: The 2024 Ticketmaster–Snowflake Breach*

In May 2024, threat actor UNC5537 (ShinyHunters) gained unauthorized access to Ticketmaster's customer data hosted in a Snowflake cloud data warehouse environment between April 14 and May 18, 2024. A joint investigation by Snowflake, CrowdStrike, and Mandiant confirmed no platform-level vulnerability or misconfiguration was involved [23]; instead, access was obtained entirely through stolen credentials against customer-managed tenant accounts. UNC5537 leveraged credentials harvested by infostealer malware families including Vidar, RisePro, Redline, Lumma, and MetaStealer [24], some of which had remained valid since 2020 with no rotation. The affected tenants lacked MFA enforcement, a control explicitly under customer ownership per Snowflake's shared responsibility model [23], along with conditional access policies restricting authentication by geography, IP range, or device compliance. The breach extended to over 165 Snowflake customer tenants, including AT&T, Santander Bank, Advance Auto Parts, and Neiman Marcus [24], confirming a systemic cross-enterprise control gap. Approximately 560 million Ticketmaster customer records (~1.3 TB) were exfiltrated, including names, contact details, encrypted payment card data, and purchase histories [24]. ShinyHunters listed the dataset for $500,000 USD [25], and a class action lawsuit followed on May 29, 2024. Snowflake subsequently mandated MFA by default for all new accounts in October 2024 [24].

Analyzed through the proposed framework, the incident reflects compounding failures across all four control domains: TPRM processes failed to identify MFA as a customer-owned residual risk requiring formal acceptance, underscoring that passing a TPRM assessment does not constitute production readiness; cybersecurity implementation reviews were absent, leaving tenant misconfigurations undetected; IAM governance permitted multi-year credential persistence without rotation or periodic access reviews; and DR/incident response when the incident was detected, the enterprise's documented ability to isolate the compromised tenant, revoke active integrations, implement emergency access restrictions, and execute recovery procedures that should have been pre-validated through tabletop or live exercises could have significantly minimized the impact radius. The absence of such rehearsed procedures likely constrained the effectiveness of the response.

## IV. THE SAAS ONBOARDING LIFECYCLE FRAMEWORK

### *A. Framework Design Principles*

The proposed framework is grounded in five foundational design principles. *(1) Control Sequencing:* Framework activities must occur in dependency order. TPRM risk rating must precede architecture approval. IAM design must precede production access provisioning. The framework encodes these dependencies as formal gates rather than advisory guidance. *(2) Pre-Contract versus Post-Contract Separation:* The framework maintains an intentional and auditable distinction between vendor risk assessment (pre-contract) and implementation control validation (post-contract). This separation preserves risk-based vendor decisioning integrity and ensures strong pre-production control enforcement. *(3) Shared Responsibility Explicitness:* Every control activity must be classified as vendor-owned, enterprise-owned, or jointly-owned. Ambiguity in shared responsibility classification is a primary driver of control gaps and a recurring finding in regulatory audits. *(4) Auditability:* Each lifecycle stage must produce documented evidence artifacts sufficient for regulatory examination, internal audit review, and external assurance. The framework specifies required evidence artifacts at each stage. *(5) Governance Continuity:* The framework does not terminate at go-live. Production deployment initiates a post-production governance phase with defined monitoring cadences, re-assessment triggers, and lifecycle events. SaaS governance is a continuous discipline, not a pre-deployment event.

### *B. The Six-Stage Onboarding Lifecycle*

The framework organizes SaaS onboarding into six sequential stages spanning pre-contract, post-contract pre-production, and continuous post-production phases. Figure 1 illustrates the Six-Stage Onboarding Lifecycle, and subsequent content within this section summarizes all six stages, their primary owners, key activities and required outputs.

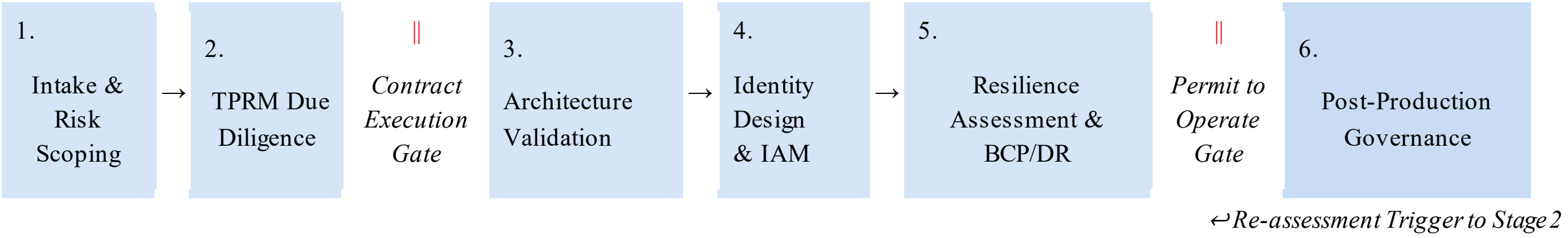

*Fig. 1. Six-Stage SaaS Onboarding Lifecycle. Pre-contract stages (1–2) separated from post-contract stages (3–5) by contract execution gate. Stage 6 represents continuous governance with reassessment triggers returning to Stage 2.*

### *C. Stage 1: Intake and Risk Scoping*

The onboarding process initiates when a business unit submits a Risk Assessment Questionnaire (RAQ) through the eGRC (Enterprise Governance, Risk Management, and Compliance) system. The RAQ captures information necessary to perform an inherent risk rating, which determines due diligence depth and control domains to engage.

Key RAQ scoping dimensions include: the nature and sensitivity classification of data to be processed or stored; estimated volume of records; replaceability index (time required to replace the vendor); countries of data processing and storage; degree of regulatory risk; extent of subcontractor delegation; and presence of internet-facing or cloud-hosted components. These dimensions are aggregated into an inherent risk rating across five tiers: Critical High, High, Medium, Low, and Nominal. Each tier carries a defined monitoring cadence and due diligence scope.

Cybersecurity Architecture engagement is mandatory regardless of TPRM risk tier when: a new SaaS platform is being adopted; new IdP or federation configurations are introduced; regulated, sensitive, or classified data is processed; cross-tenant integrations with external parties are established; or internet-facing components are deployed. Stage 1 outputs: formal risk tier classification in the GRC system; TPRM engagement confirmation; and Cybersecurity Architecture engagement confirmation where applicable.

### *D. Stage 2: TPRM Due Diligence and Vendor Assessment*

Stage 2 is the pre-contract vendor-level assessment phase, owned by the TPRM/TRM function. Its governing question: "Do we trust this SaaS provider's security and operational posture sufficiently to execute a contract?"

The cybersecurity TPRM assessment evaluates: the vendor's information security governance framework; data protection commitments and mechanisms; identity and access control model; vulnerability management practices; incident response capabilities; breach notification obligations and timelines; and subcontractor security controls. Primary evidence sources include

SOC 2 Type II reports, ISO/IEC 27001:2022 certification, penetration testing executive summaries, and vendor-completed due diligence questionnaires covering cybersecurity, BCP/DR, privacy, compliance, insurance, and financial health.

The BCP/DR TPRM assessment evaluates: vendor formal BCP and DR plans; stated RTO and RPO targets; primary and alternate processing locations; resilience testing cadence and results; ISO 22301:2019 certification status; and contractual willingness to participate in joint exercises.

Assessment outputs include: a cybersecurity risk rating (Low/Moderate/High/Critical); a BCP/DR risk rating; identification of required contractual clauses (incident notification timelines, audit rights, data protection obligations, resilience commitments, data return and destruction upon termination); and a formal risk acceptance or rejection decision documented in the eGRC system.

The critical principle governing the boundary between Stage 2 and Stage 3: passing TPRM does not mean the SaaS application is production-ready. TPRM establishes that the vendor's security program meets the enterprise's acceptable risk threshold. Post-contract reviews establish whether the enterprise's specific use of the vendor's platform is securely configured and operationally sound.

Figure 2 describes TPRM Due Diligence Process Flow: The sequential flow proceeds as follows - RAQ Submission → Inherent Risk Rating → TPRM Engagement Triggered → Vendor DDQ(Due Diligence Questionnaire) Issuance → SOC 2 / ISO Review → Cyber Risk Rating → BCP/DR Rating → Risk Acceptance Decision → Contract Execution Gate. Pre-contract stages (1–2) are separated from post-contract stages by a formal contract execution gate. The gate requires documented approval from the TPRM function and records the risk tier, cyber rating, and BCP/DR rating in the enterprise GRC system before contract signature proceeds. Risk ratings drive both contractual clause requirements and the control intensity applied in Stages 3–5.

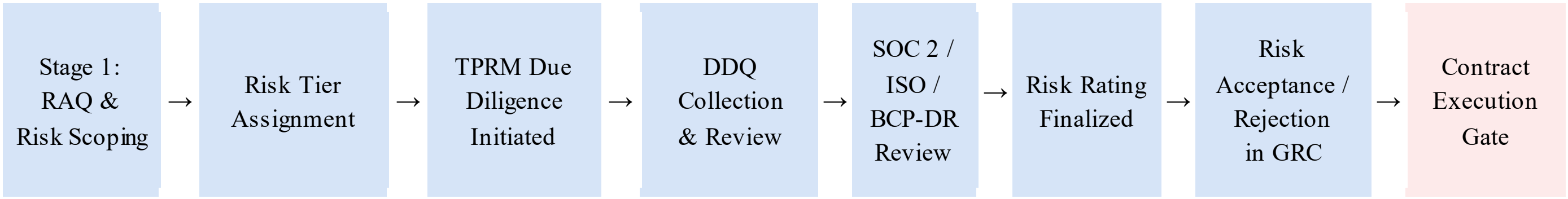

*Fig. 2. TPRM Due Diligence Process Flow - sequential gate model from RAQ submission through contract execution. Pre-contract stages 1–2 are formally separated from post-contract stages by the contract execution gate.*

*E. Stage 3: Architecture Validation and Cybersecurity Review*

Stage 3 is the post-contract, pre-production implementation validation phase, owned by the Cybersecurity Architecture and Controls Assessment functions. Its governing question: "Is this SaaS service securely configured and operating in alignment with enterprise security standards before go-live?"

Unlike Stage 2, which evaluates vendor-level claims and attestations, Stage 3 validates actual tenant configuration through direct review of provisioned settings, architecture documentation, and technical evidence. Review scope encompasses: SaaS consumption model validation; data ingress, egress, storage, and exposure path mapping; secure configuration baseline review; encryption at rest (with preference for Customer-Managed Keys/BYOK where supported) and in transit (TLS 1.2 minimum); logging, monitoring, and audit access capabilities; data handling alignment with enterprise data classification and retention standards; identity federation configuration, MFA enforcement, RBAC structure, and least-privilege validation; API and integration security; and shared responsibility boundary documentation.

A critical output is the explicit mapping of four trust boundaries: (1) the Enterprise User Environment (enterprise endpoints, network access, enterprise IdP); (2) the Internet/External Network Boundary (access path from enterprise to SaaS); (3) the Vendor SaaS Tenant Boundary (enterprise-configurable tenant layer); and (4) the Vendor Internal Platform Boundary (vendor-managed infrastructure inaccessible to the enterprise).

Stage 3 formally produces: a Requirements Specification Document (RSD) or control baseline; a gap list with remediation plans and owners; a Permit to Operate (PTO) approval or conditional restricted enablement with defined remediation timelines; and SaaS Security Posture Management (SSPM) enrollment confirmation for critical platforms.

*F. Stage 4: Identity Design and IAM Integration*

Stage 4 is the post-contract, pre-production identity governance phase, owned by the Enterprise IAM function. Its governing question: "Are all users and administrators securely authenticated and governed through enterprise-controlled identity infrastructure?"

The foundational requirement: all users and administrators must authenticate to the SaaS platform via the centralized enterprise Identity Provider. Direct vendor-native authentication is prohibited for all standard users and must be limited to documented break-glass emergency access accounts with defined monitoring and review procedures. MFA is mandatory for all privileged and administrative access; this is an architectural requirement validated through IAM governance evidence, not a configuration recommendation.

RBAC design must define roles explicitly based on job function, applying the principle of least privilege. Default vendor roles conferring excessive permissions must be replaced or constrained. Shared or group accounts are prohibited without a documented exception. PAM controls must be applied to all administrative accounts, with vaulted credentials, session monitoring, and just-in-time access provisioning where technically feasible.

Formal onboarding and offboarding processes must be defined, documented, and aligned to HR workflow triggers to ensure timely revocation of access upon employee termination or role change. Access reviews must be conducted periodically, with frequency determined by the platform risk tier.

Stage 4 evidence artifacts: IAM architecture documentation with federation design; IdP federation configuration attestation; access review records; PAM enrollment confirmation; and onboarding/offboarding procedure documentation.

### *G. Stage 5: Resilience Assessment and BCP/DR Planning*

Stage 5 is the post-contract, pre-production operational resilience validation phase, required for all Critical High and High risk-tier SaaS platforms. Its governing question: "Can this SaaS service support enterprise recovery needs when an actual disruption occurs?", which is materially different from Stage 2's question of "Does this vendor appear capable of recovering its service in theory?"

Stage 5 validates actual operational resilience by: aligning vendor-stated SLA RTO/RPO commitments with enterprise business recovery objectives; mapping dependencies outside the vendor's control sphere (identity infrastructure availability, network access path redundancy, dependent integration availability); establishing clarity on roles and responsibilities during an actual disruption; and validating communication and escalation paths between enterprise and vendor incident response teams.

Activities include: service-level BCP/DR planning documentation; tabletop exercises around realistic disruption scenarios; participation in enterprise or regulatory resilience exercises where the SaaS service is a dependency; and documentation of results, identified gaps, and remediation commitments with owners and timelines.

Stage 5 produces: confirmed service-level recovery expectations as a service BCP; tabletop exercise results and lessons learned; identified recovery gaps with corrective action plans; and evidence artifacts for regulatory and audit review.

### *H. Stage 6: Post-Production Governance and Ongoing Monitoring*

Stage 6 is the continuous governance phase that prevents control drift and maintains the security and resilience posture established through Stages 1–5. It is jointly owned across TPRM, Cybersecurity, IAM, and Service Owner (Business Unit) functions.

TPRM re-assessment cadences are tiered by risk: Critical High and High risk-tier vendors re-assessed annually; Medium risk-tier vendors every two years; Low risk-tier vendors every three years; Nominal risk-tier vendors on an event-driven basis. SSPM continuous monitoring is required for all Critical High and High risk-tier SaaS platforms, with findings triaged and remediated through a defined workflow.

Periodic access recertification and role reviews are conducted on a schedule commensurate with the platform risk tier. SLA and KPI monitoring are conducted with the vendor at minimum quarterly frequency for High and Critical High risk-tier platforms. Contract renewal reviews provide a structured opportunity to update terms and conditions. Incident reporting obligations require business units to identify, analyze, document, and respond to third-party incidents that result in actual or potential financial loss, service disruption, or reputational damage.

## V. CROSS-DOMAIN CONTROL MAPPING

When TPRM, cybersecurity, IAM, and DR teams operate independently without a shared sequencing model or dependency map, specific and predictable failure modes emerge. TPRM teams may approve a vendor while cybersecurity teams apply a higher-stringency assessment that contradicts the risk acceptance, with neither team being aware of the conflict. IAM designs may be finalized without knowledge of cybersecurity architecture requirements, resulting in configurations that satisfy IAM standards but violate cybersecurity controls. DR plans may assume identity infrastructure availability without validation that the enterprise IdP can recover within the SaaS platform RTO. Across all these failure modes, no single function maintains visibility of the cumulative control posture.

### *A. Cross-Domain Control Mapping*

In this framework, "cross-domain" refers to the intersection of the four interdependent control domains governing SaaS onboarding. Each domain produces controls, evidence artifacts, and risk decisions that directly affect the others; cross-domain mapping makes these dependencies explicit, surfacing the failure modes that arise when any one domain executes in isolation.

Table II presents the cross-domain control mapping for SaaS onboarding, identifying control activities by domain, whether they apply pre- or post-contract, the required evidence artifact, and the common failure mode observed when controls are executed in siloed fashion.

TABLE II.

Cross-Domain Control Mapping for SaaS Onboarding

| Domain | Control Activity | Pre-Contract (TPRM) | Post-Contract (Implementation) | Evidence Artifact | Common Failure Mode (Siloed) |
|---|---|---|---|---|---|
| Cyber security | Risk Rating | Vendor cyber risk rated via TPRM DDQ and SOC 2 | Tenant-level configuration rated via architecture review | TPRM Cyber Risk Rating; RSD Control Baseline | Vendor approved at TPRM; tenant misconfigured in production |
| | MFA Enforcement | Vendor MFA capability and default posture assessed via DDQ | MFA enforcement validated on enterprise tenant; confirmed in IAM evidence | IAM architecture doc; federation attestation; SSPM report | MFA capability confirmed at vendor level; not enforced on tenant |
| | Encryption | Vendor encryption capabilities reviewed (at-rest/in-transit) | CMK/BYOK configuration validated; TLS version confirmed on tenant | SOC 2 encryption section; TPRM review; contractual validation | Vendor encryption attested; enterprise tenant uses vendor-managed keys only |
| | Audit Logging | Vendor logging capability and log access policy assessed | Enterprise access to tenant audit logs confirmed; SIEM integration validated | Logging architecture documentation; SIEM integration evidence | Vendor logs exist but enterprise cannot access them; incident response is blind |
| | Incident Notification | Contractual notification SLA negotiated (e.g., 72 hours) | Notification workflow tested; escalation contacts confirmed | Contract clause; tested escalation path documentation | Contract clause agreed; no tested notification path; first contact during real incident |
| | SSPM Monitoring | Vendor SSPM compatibility and API access assessed | SSPM tool enrolled; continuous monitoring baseline established | SSPM enrollment record; initial baseline report | SSPM not enrolled at go-live; configuration drift undetected post-production |
| TPRM | Vendor Selection | Inherent risk rating determines due diligence scope | N/A (pre-contract activity) | RAQ; inherent risk rating; GRC record | Risk tier assigned incorrectly; insufficient due diligence for critical vendor |
| | Due Diligence Questionnaires | Cyber, BCP/DR, Privacy, Compliance, Insurance DDQs completed | DDQ responses validated against implementation; gaps escalated | Completed DDQs; vendor responses; scoring records | Cyber and BCP DDQs sent separately by different teams; vendor receives duplicates |
| | Contractual Clauses | Required clauses identified; negotiated into contract | Clauses verified as executed in signed agreement | Signed contract with required exhibits; TPRM clause checklist | Clauses omitted in final contract; discovered only at re-assessment |
| | Ongoing Monitoring | Monitoring cadence established per risk tier | SSPM findings feed TPRM re-assessment evidence | Re-assessment reports; SSPM finding summaries | TPRM re-assessment relies solely on vendor attestations; SSPM findings not integrated |
| IAM | Federated Identity | Vendor IdP federation support and SAML/OIDC capability assessed | Federation configured; tested; native accounts disabled | IdP federation configuration; test evidence; native account inventory | Federation not completed at go-live; native accounts remain active permanently |
| | RBAC Design | Vendor RBAC model and default role structure assessed | Enterprise-defined roles documented; least-privilege applied; default roles replaced | RBAC role definition documentation; access review records | Default vendor roles adopted; broad permissions persist; no role review conducted |
| | PAM Enrollment | Vendor support for PAM integration assessed | Administrative accounts enrolled in PAM; session monitoring confirmed | PAM enrollment records; administrative account inventory | Administrative accounts managed outside PAM; credentials exposed in shared tools |
| | Access Reviews | Vendor access review functionality assessed | Periodic review schedule established; first review conducted within 90 days of go-live | Access review records; certification sign-offs | Access reviews not scheduled; accounts accumulate; orphaned accounts go undetected |
| | Offboarding | Vendor API/SCIM offboarding capability assessed | Offboarding workflow integrated with HR; tested end-to-end | Offboarding procedure documentation; test evidence | Manual offboarding; delay between termination and access revocation |

| Domain | Control Activity | Pre-Contract (TPRM) | Post-Contract (Implementation) | Evidence Artifact | Common Failure Mode (Siloed) |
|---|---|---|---|---|---|
| BCP/DR | RTO/RPO Assessment | Vendor RTO/RPO commitments reviewed and contractualized | Enterprise business recovery objectives validated against vendor SLA | BCP/DR risk rating; contract SLA; service BCP documentation | Vendor RTO accepted without validating enterprise-side recovery dependencies |
| | Joint Exercises | Vendor willingness and availability for joint exercises confirmed | Tabletop exercise conducted; results documented; gaps remediated | Tabletop exercise results; lessons learned; gap remediation records | Exercise conducted at vendor level only; enterprise recovery procedures untested |
| | Dependency Mapping | Vendor subcontractor and infrastructure dependencies assessed | Enterprise-side dependencies mapped: IdP, network, integrations | Dependency map; service BCP documentation | IdP dependency not mapped; enterprise unable to authenticate during SaaS DR event |
| | Recovery Sequencing | Vendor recovery sequence and communication protocol reviewed | Enterprise recovery sequence documented and tested for this specific SaaS service | Recovery runbook; tabletop exercise evidence | Generic DR plan applied; SaaS-specific recovery sequence undefined; incident response improvised |

*DDQ = Due Diligence Questionnaire; RSD = Requirements Specification Document; SIEM = Security Information and Event Management; SSPM = SaaS Security Posture Management; BCP = Business Continuity Plan.*

## *B. Sequencing Dependencies*

The framework defines a mandatory sequencing model in which each stage gate produces outputs that are required inputs for subsequent stages. The sequential pipeline is described in Figure 3.

Dependency Gate Structure - Chain of Actions**:** (1) Inherent Risk Scoping Complete → (2) TPRM Due Diligence Complete → (3) Contract Executed → (4) Architecture Validation Complete → (5) IAM Design Finalized → (6) DR Dependency Mapping Complete → (7) Permit to Operate (PTO) Issued → Production Data Enablement. Each arrow represents a mandatory dependency: downstream gates cannot open until upstream gates are formally closed with documented evidence. A Policy Exception bypass node is shown with mandatory SVP/EVP approval sign-off requirement. Exception records are retained permanently in the GRC system as audit findings with assigned remediation owners and target completion dates. No gate may be bypassed through informal process escalation.

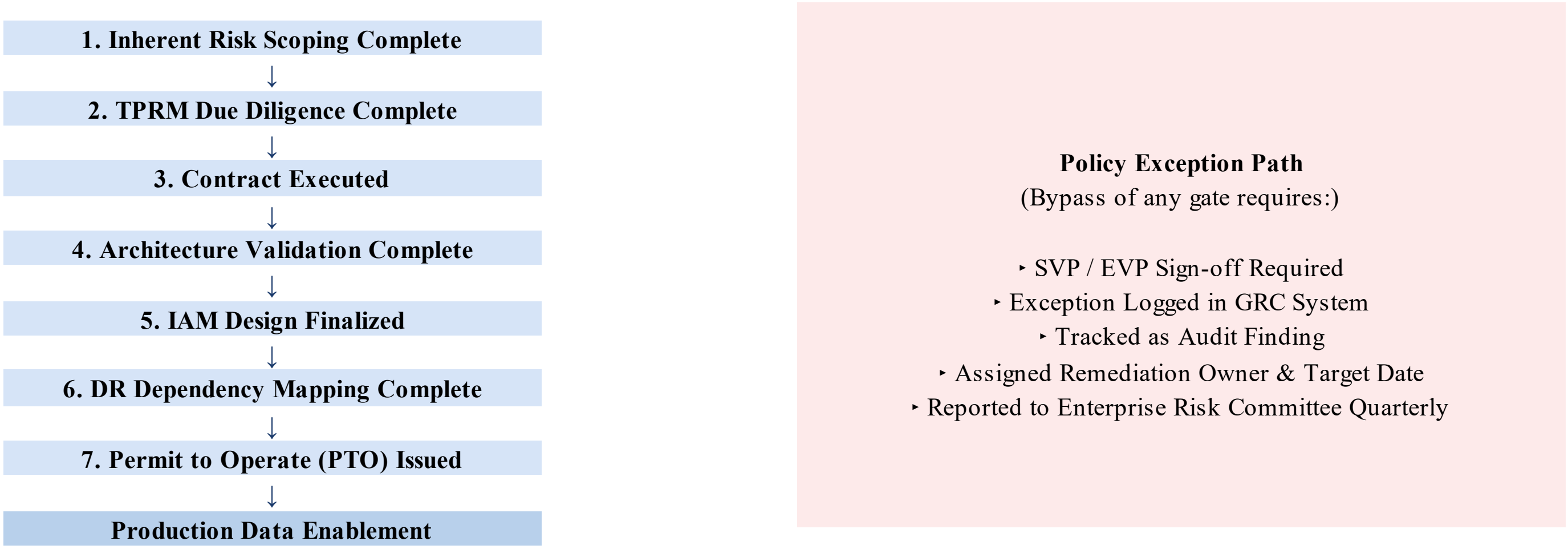

*Fig. 3. Mandatory Sequencing Dependency Chain of Actions. Each gate requires formal evidence before downstream gates open. Policy exceptions require SVP/EVP approval and are recorded as permanent GRC audit findings.*

The rationale for each dependency ordering is as follows: (1) Inherent risk scoping must precede TPRM due diligence because the risk tier determines the scope and depth of due diligence required; a High-risk platform requires more extensive DDQs than a Nominal-risk platform. (2) TPRM approval must precede contract execution because contract execution without TPRM approval constitutes a policy violation that bypasses the enterprise's primary vendor risk gate. (3) Contract execution must precede architecture and cybersecurity review because the post-contract review validates the enterprise's specific tenant configuration, which requires a provisioned and contracted environment. (4) Architecture validation must precede IAM design finalization because architecture review may identify trust boundary requirements, data classification constraints, or integration restrictions that directly affect IAM role design and federation configuration. (5) IAM design must precede DR dependency mapping because identity infrastructure availability is a primary enterprise-side dependency during a SaaS DR event, and this dependency cannot be accurately mapped until the IAM architecture is finalized. (6) All pre-production controls must be validated before the Permit to Operate because the PTO is the formal, accountable evidence artifact authorizing production data enablement; issuing PTO with open control gaps creates documented residual risk. (7) PTO must precede production data enablement because enabling production data flows without PTO authorization constitutes a bypass of the enterprise's pre-production security gate.

### *C. Key Architectural Risk Conditions*

The framework identifies eight architectural risk conditions representing high-priority findings in SaaS security assessments: *(1) Over-privileged administrative roles:* The most common finding in enterprise SaaS governance reviews. Administrative roles assigned for convenience expand the blast radius of any credential compromise event. *(2) Excessive or unclassified data ingestion:* When the volume and sensitivity classification of data introduced to a SaaS platform is not formally documented and approved, the enterprise cannot accurately assess its breach exposure. *(3) Inadequate logging or restricted audit access:* Where a SaaS vendor does not provide enterprise access to tenant audit logs, this must be treated as a high-severity control gap that prevents forensic analysis and regulatory compliance. *(4) Weak vendor token or API key governance:* Unrotated API keys are a primary persistence mechanism for threat actors following initial access. *(5) Dependency on vendor incident response without enterprise validation:* Enterprises must maintain their own incident response capabilities and communication protocols, not rely exclusively on vendor-initiated notification and response. *(6) Data stored without customer-managed encryption keys:* Where CMK or BYOK is technically supported, it must be implemented for regulated data. Where not supported, this gap must be documented as a residual risk. *(7) Undefined data retention and deletion models:* Data retained beyond operational need represents unnecessary exposure. Data not formally deleted upon contract termination may persist in violation of regulatory obligations. *(8) Undefined shared responsibility model:* The root cause of the most severe and persistent SaaS security failures, including the Ticketmaster–Snowflake breach. The shared responsibility model must be formally documented, contractually referenced, and reviewed at each TPRM re-assessment cycle.

## VI. DESIGN PATTERNS FOR SECURE SAAS ONBOARDING

The lifecycle framework and cross-domain control mapping presented in Sections IV and V establish what must be done and in what sequence across the SaaS onboarding lifecycle. This section translates that structural guidance into six reusable design patterns as prescriptive, implementation-level constructs that specify how key control requirements should be designed and operationalized in practice. Each pattern addresses one or more of the architectural risk conditions identified in Section V and is expressed in terms of context, design intent, associated anti-patterns, and required control evidence. Collectively, the six patterns constitute a reusable design library applicable across SaaS deployment models, platform types, and regulated industry contexts. They are intended to support architecture teams during Stages 3 through 5 of the onboarding lifecycle and to inform ongoing governance activities in Stage 6.

### *A. Pattern 1: Federated Identity Integration*

*Context:* SaaS platforms supporting SAML 2.0 or OpenID Connect/OAuth 2.0 federation with enterprise Identity Providers. *Design Intent:* All authentication flows for enterprise users and administrators must be routed through the centralized enterprise IdP (e.g., Azure Entra ID, Okta, Ping Identity). Vendor-native identity stores must be disabled upon federation activation. MFA must be enforced at the IdP level, ensuring all SaaS platform sessions are subject to enterprise MFA policy regardless of whether the SaaS platform itself enforces MFA. This approach centralizes authentication governance, enables session policy enforcement (conditional access, device compliance), and ensures that offboarding at the enterprise IdP level immediately terminates SaaS platform access without requiring separate action in each SaaS tenant. *Anti-Pattern:* Allowing vendor-managed local accounts for administrative users during onboarding as a temporary measure. In practice, these accounts become permanent, accumulate over time, and are a recurring finding in TPRM and external audit reviews. The Ticketmaster–Snowflake breach involved credentials that persisted and remained valid since 2020 due to exactly this pattern. *Control Evidence:* IAM architecture documentation with federation design diagram; IdP federation configuration attestation; native account inventory confirming disablement; access review records.

### *B. Pattern 2: Least-Privilege Role Architecture*

*Context:* Any SaaS platform supporting configurable Role-Based Access Control. *Design Intent:* Roles must be explicitly defined based on enterprise job function requirements, not adopted from vendor default role templates. Vendor default roles conferring broad or administrative permissions must be reviewed, constrained, or replaced. Shared accounts and group accounts are prohibited without a formal, documented exception. Administrative roles must be segregated from standard user roles and subject to PAM controls. The blast radius of any individual credential compromise is bounded by the scope of the compromised account's assigned role, making least-privilege enforcement a direct mitigant against credential-based attacks. *Ticketmaster Relevance:* The absence of proper role governance meant that once credentials were compromised, the threat actor had access to the full scope of data accessible to the compromised account enabling exfiltration of 560 million records. *Control Evidence:* Role definition documentation with job-function mapping; access certification records; exception tracking for any shared accounts; PAM administrative account inventory.

### C. Pattern 3: Encryption and Key Governance

*Context:* SaaS platforms supporting BYOK or CMK capabilities, applicable to any platform processing regulated or sensitive data. *Design Intent:* Enterprise data at rest within SaaS platforms must be protected using enterprise-controlled encryption keys where technically supported. Key ownership, custody, and rotation schedules must remain under enterprise governance, stored in an enterprise-managed Hardware Security Module (HSM) or Key Management Service (KMS). Key rotation must follow a defined schedule. Data in transit must be encrypted using TLS 1.2 or higher, validated during the architecture review. *Risk Condition:* Data stored without enterprise-controlled encryption creates a dependency on the SaaS vendor's key management practices and significantly increases exposure in multi-tenant breach scenarios. If a vendor manages all tenant encryption keys and vendor key management is compromised or if a government order compels key disclosure, then the enterprise has no independent cryptographic protection. *Control Evidence:* CMK/BYOK configuration documentation; key rotation schedule; TPRM review of vendor encryption architecture; SOC 2 encryption controls section review; contractual validation of key management obligations.

### D. Pattern 4: Secure API and Integration Governance

*Context:* SaaS platforms with API integrations connecting to enterprise systems, data pipelines, or other SaaS services. *Design Intent:* API tokens and service account credentials must be scoped to the minimum permissions required for the documented integration purpose. Token rotation must be enforced on a defined schedule. Service account inventories must be maintained and included in periodic access reviews. Integration authentication mechanisms must be documented as security-relevant design elements in architecture documentation. API gateway or proxy architectures should be employed where feasible to centralize integration governance and monitoring. *Risk Condition:* Unrotated API tokens are a primary persistence mechanism for threat actors following initial access. Once an API token is compromised, an attacker can maintain access even if the primary user credential is remediated, because API tokens often operate independently of the user authentication layer. *Control Evidence:* API integration inventory documentation; token rotation policy and evidence of compliance; service account review records; integration architecture diagram reviewed in Stage 3.

### E. Pattern 5: Shared-Responsibility DR Design

*Context:* SaaS platforms used for critical business functions classified at the High or Critical High risk tier. *Design Intent:* A formal, documented shared responsibility matrix must explicitly identify which recovery controls are vendor-owned and which are enterprise-owned for each relevant failure scenario. Enterprise recovery objectives must be validated against vendor contractual SLA commitments through operational tabletop exercises, not solely through vendor attestation or SOC 2 review. Dependencies outside the vendor's control sphere must be explicitly identified, mapped, and tested: enterprise IdP availability; network connectivity path redundancy; and upstream/downstream integration availability. *Anti-Pattern:* Treating the TPRM BCP/DR questionnaire response or SOC 2 availability attestation as equivalent to proof of operational resilience for a specific enterprise deployment. Vendor resilience capability is a necessary but not sufficient condition for enterprise operational resilience. *Control Evidence:* Service-level BCP documentation with shared responsibility matrix; tabletop exercise results; dependency map; recovery sequence runbook; regulatory exercise participation evidence.

### F. Pattern 6: Continuous SaaS Security Posture Management (SSPM)

*Context:* Critical High and High risk-tier SaaS platforms post-production. *Design Intent:* SSPM tooling continuously monitors SaaS tenant configuration for drift from the approved baseline established during Stage 3, policy violations, over-privileged accounts, inactive user accounts, and deviations from security benchmarks. Findings are triaged through a defined workflow with severity-based remediation timelines. Critical SSPM finding categories include: MFA enforcement drift; new native accounts outside the federated identity model; RBAC role assignment changes; data sharing and export policy changes; and integration permission scope expansions. SSPM findings are formally integrated into the TPRM re-assessment evidence package, providing the TPRM team with an objective, continuous view of the enterprise's actual security posture on the platform rather than relying solely on vendor-provided attestations at re-assessment time. *Control Evidence:* SSPM enrollment record; continuous monitoring dashboard access; finding trend reports; remediation tracking records; integration with TPRM re-assessment evidence package.

## VII. REFERENCE CHECKLIST FOR SAAS ONBOARDING

The SaaS Onboarding Reference Checklist is a structured governance instrument organized by lifecycle stage, designed for use by onboarding teams, control owners, internal audit functions and regulatory examiners. Each item maps directly to a control activity defined in the six-stage framework described in Section IV, and specifies the responsible owner and required evidence artifact, enabling traceable accountability across all four control domains - TPRM, cybersecurity, IAM, and DR.

The checklist comprises control items spanning six lifecycle stages. Stage 1 items govern intake completeness, risk scoping, and regulatory applicability. Stage 2 items address vendor due diligence and third-party attestation review. Stage 3 items cover architecture validation, connectivity design, and Permit to Operate issuance. Stage 4 items enforce federated identity, MFA, and least-privilege access provisioning. Stage 5 items assess vendor and enterprise resilience, DR plan alignment, and tabletop exercise execution. Stage 6 items govern post-production monitoring, SSPM coverage, periodic reassessment, and ongoing risk reporting. Across all stages, each item specifies a single designated control owner to prevent the accountability gaps that arise from siloed, uncoordinated reviews.

The checklist is intended to be maintained in the enterprise GRC system as a living record, with evidence artifacts linked to each item at the point of completion. Open items and policy exceptions must be formally tracked with assigned owners and target remediation dates and reported to the enterprise risk committee as part of the Stage 6 governance cadences described in Section IV.H. This approach supports both internal audit reviews and regulatory examination preparation by providing a single, traceable record of control execution across the full onboarding lifecycle. The complete checklist, covering control items across all six stages, is provided in Table II.

TABLE II.

SaaS Onboarding Reference Checklist (Organized by Lifecycle Stage)

| Ref. | Control Item | Owner | Evidence Artifact |
|---|---|---|---|
| **STAGE 1 - INTAKE & RISK SCOPING** | | | |
| 1.1 | Risk Assessment Questionnaire (RAQ) completed and submitted via GRC system | Business Unit | Completed RAQ in GRC system |
| 1.2 | Data sensitivity classification assigned to all data types to be processed | Business Unit / Data Owner | Data classification record |
| 1.3 | Inherent risk tier determined (Critical High / High / Medium / Low / Nominal) | TPRM / Risk Management | GRC risk tier record |
| 1.4 | TPRM engagement formally triggered based on risk tier threshold | Business Unit | TPRM engagement confirmation in GRC |
| 1.5 | Cybersecurity Architecture engagement triggered (mandatory for new SaaS platforms) | Business Unit | Cybersecurity Architecture engagement confirmation |
| 1.6 | Countries of data processing and storage identified and documented | Business Unit | Data residency documentation in RAQ |
| 1.7 | Estimated user population and access scope defined | Business Unit | User scope documentation in RAQ |
| 1.8 | Subcontractor and fourth-party delegation scope identified | TPRM | Subcontractor scope in RAQ |
| 1.9 | Regulatory applicability assessed (HIPAA, GDPR, DORA, SOX, FFIEC, etc.) | Compliance / Business Unit | Regulatory applicability determination |
| 1.10 | Replaceability index and vendor concentration risk documented | TPRM | Concentration risk assessment in GRC |
| **STAGE 2 - TPRM DUE DILIGENCE & VENDOR ASSESSMENT** | | | |
| 2.1 | Vendor cybersecurity due diligence questionnaire (DDQ) issued and completed | TPRM | Completed cyber DDQ on file in GRC |
| 2.2 | Vendor BCP/DR due diligence questionnaire issued and completed | TPRM | Completed BCP/DR DDQ on file in GRC |
| 2.3 | SOC 2 Type II report reviewed; trust services criteria mapped to risk | TPRM / Cybersecurity | SOC 2 review memo |
| 2.4 | ISO/IEC 27001:2022 certification validated (where applicable) | TPRM | ISO 27001 certificate copy; validity confirmed |
| 2.5 | ISO 22301:2019 certification or equivalent BCP standard validated | TPRM | ISO 22301 certificate or equivalent documentation |
| 2.6 | Vendor cybersecurity risk rating formally determined and recorded | TPRM | Cyber risk rating record in GRC |
| 2.7 | Vendor BCP/DR risk rating formally determined and recorded | TPRM | BCP/DR risk rating record in GRC |
| 2.8 | Required contractual clauses negotiated: breach notification timeline, audit rights, data protection, DR participation, data return/destruction | TPRM / Legal | Signed contract with required clause exhibit |
| 2.9 | Shared Responsibility Model (SRM) documentation obtained and reviewed | TPRM / Cybersecurity | SRM documentation on file |
| 2.10 | Formal risk acceptance or rejection decision documented in GRC; contract execution gate cleared | TPRM / Risk Officer | Risk acceptance record in GRC; contract signed |

| Ref. | Control Item | Owner | Evidence Artifact |
|---|---|---|---|
| **STAGE 3 - ARCHITECTURE VALIDATION & CYBERSECURITY REVIEW** | | | |
| 3.1 | SaaS consumption model and tenant architecture documented | Cybersecurity Architecture | Architecture documentation / RSD |
| 3.2 | Data ingress, egress, storage, and exposure paths mapped and classified | Cybersecurity Architecture | Data flow diagram in RSD |
| 3.3 | Four trust boundaries explicitly documented: Enterprise User, Internet Boundary, Vendor Tenant, Vendor Platform | Cybersecurity Architecture | Trust boundary map in RSD |
| 3.4 | Encryption at rest validated; CMK/BYOK implemented where technically supported | Cybersecurity Architecture | CMK/BYOK configuration record |
| 3.5 | Encryption in transit validated; TLS 1.2 minimum confirmed | Cybersecurity Architecture | TLS configuration validation record |
| 3.6 | Tenant audit logging confirmed; SIEM integration validated | Cybersecurity / SIEM Team | SIEM integration confirmation; log completeness report |
| 3.7 | Shared responsibility model formally documented and agreed with vendor | Cybersecurity Architecture | Signed SRM exhibit in contract |
| 3.8 | Control gap list produced with remediation plans, owners, and timelines | Cybersecurity Architecture | Gap list and remediation plan in GRC |
| 3.9 | SSPM enrollment completed for Critical High and High risk-tier platforms | Cybersecurity / SSPM Team | SSPM enrollment record; baseline configuration export |
| 3.10 | Permit to Operate (PTO) issued or conditional restricted enablement approved with defined remediation timeline | Cybersecurity Architecture | PTO approval record in GRC |
| **STAGE 4 - IDENTITY DESIGN & IAM INTEGRATION** | | | |
| 4.1 | Enterprise IdP federation configured (SAML 2.0 or OIDC/OAuth 2.0) | Enterprise IAM | IdP federation configuration attestation |
| 4.2 | MFA enforced at enterprise IdP level for all SaaS platform users | Enterprise IAM | MFA enforcement configuration evidence; IAM architecture doc |
| 4.3 | Vendor-native local accounts inventoried and disabled upon federation activation | Enterprise IAM | Native account inventory; disablement confirmation |
| 4.4 | Break-glass emergency access accounts defined, documented, monitored, and reviewed | Enterprise IAM | Break-glass account documentation; monitoring confirmation |
| 4.5 | RBAC role model defined based on enterprise job function; default vendor roles reviewed and constrained | Enterprise IAM | Role definition documentation with job-function mapping |
| 4.6 | Least-privilege access validated; no over-privileged default role assignments in use | Enterprise IAM | Access certification records |
| 4.7 | Administrative accounts enrolled in PAM; vaulted credentials and session monitoring enabled | Enterprise IAM / PAM Team | PAM administrative account inventory; enrollment confirmation |
| 4.8 | Shared or group accounts prohibited; exceptions formally documented | Enterprise IAM | Exception tracking record (if applicable) |
| 4.9 | Onboarding process aligned to HR joiner workflow; access provisioning automated where possible | Enterprise IAM | Onboarding procedure documentation; HR integration confirmation |
| 4.10 | Offboarding process aligned to HR termination workflow; access revocation validated within defined SLA | Enterprise IAM | Offboarding procedure documentation; revocation SLA validation |
| **STAGE 5 - RESILIENCE ASSESSMENT & BCP/DR PLANNING** | | | |
| 5.1 | Vendor-stated RTO and RPO documented and validated against enterprise Business Recovery Objectives | Business Unit / Enterprise Resilience | RTO/RPO alignment documentation in service BCP |
| 5.2 | Enterprise-side dependencies mapped: IdP availability, network paths, dependent integrations | Enterprise Resilience / IAM | Dependency map; service BCP |

| Ref. | Control Item | Owner | Evidence Artifact |
|---|---|---|---|
| 5.3 | Shared responsibility matrix for DR produced and agreed with vendor | Enterprise Resilience / TPRM | DR shared responsibility matrix in service BCP |
| 5.4 | Vendor escalation paths and emergency contact protocols documented | Business Unit / Enterprise Resilience | Vendor escalation contact list in service BCP |
| 5.5 | Tabletop exercise conducted for Critical High and High risk-tier platforms; realistic disruption scenarios included | Enterprise Resilience | Tabletop exercise results and lessons learned |
| 5.6 | IdP outage scenario specifically tested in tabletop (authentication unavailability during SaaS DR) | Enterprise Resilience / IAM | Tabletop results including IdP outage scenario |
| 5.7 | Integration recovery sequencing validated; upstream/downstream dependency order documented | Enterprise Resilience | Recovery sequence runbook |
| 5.8 | Identified recovery gaps documented with corrective action plans, owners, and target completion dates | Enterprise Resilience | Gap remediation plan |
| 5.9 | Regulatory resilience exercise participation confirmed (where required by DORA, FFIEC, or other mandates) | Enterprise Resilience / Compliance | Regulatory exercise participation evidence |
| 5.10 | Service-level BCP document finalized and approved; stored in enterprise BCP repository | Business Unit / Enterprise Resilience | Approved service BCP document |
| **STAGE 6 - POST-PRODUCTION GOVERNANCE & ONGOING MONITORING** | | | |
| 6.1 | TPRM re-assessment scheduled per risk tier cadence (Critical High/High: annual; Medium: every 2 years; Low: every 3 years) | TPRM | Re-assessment schedule in GRC; completed re-assessment reports |
| 6.2 | SSPM monitoring active; findings triaged and remediated per severity SLA | Cybersecurity / SSPM Team | SSPM finding reports; remediation tracking records |
| 6.3 | SSPM findings formally incorporated into TPRM re-assessment evidence package | TPRM / Cybersecurity | SSPM trend report included in re-assessment package |
| 6.4 | Periodic access recertification conducted per risk tier schedule | Enterprise IAM / Business Unit | Access certification records |
| 6.5 | RBAC role assignments reviewed; over-privileged assignments remediated | Enterprise IAM | Role review records; remediation tracking |
| 6.6 | SLA and KPI reviews conducted at minimum quarterly for High and Critical High risk-tier platforms | Business Unit / TPRM | SLA review minutes; KPI trend reports |
| 6.7 | Contract renewal review conducted; terms and conditions updated as required | TPRM / Legal | Contract renewal record; updated terms documentation |
| 6.8 | Third-party incidents reported, analyzed, and documented; impact to enterprise assessed | Business Unit / Cybersecurity | Incident log; impact assessment; remediation record |
| 6.9 | BCP/DR tabletop exercises refreshed annually for Critical High and High risk-tier platforms | Enterprise Resilience | Annual tabletop exercise records |
| 6.10 | Policy exceptions and open control gaps reported to enterprise risk committee quarterly | TPRM / Risk Officer | Quarterly risk committee reporting; exception tracking log |

## VIII. DISCUSSION

### *A. Framework Applicability Across Regulated Industries*

The framework is designed with regulatory applicability as a first-order design constraint. In financial services, the TPRM risk tier structure maps directly to enhanced due diligence requirements articulated in FFIEC IT Examination Handbook guidance and OCC Bulletin 2013-29, which requires due diligence proportionate to the risk and complexity of the third-party relationship [2], [5]. The framework's explicit evidence artifact requirements align with regulatory examination expectations for demonstrable, documented control execution.

In healthcare, the framework supports compliance with HIPAA Security Rule requirements for business associate agreements and technical safeguards governing SaaS platforms that process Protected Health Information (PHI). The Permit to Operate gate provides a documented authorization record consistent with HIPAA's authorization and access control requirements. In the European Union, the Digital Operational Resilience Act (DORA), which took effect January 17, 2025, establishes specific requirements for ICT third-party risk management, incident reporting, and operational resilience testing that map directly to Stages 2, 5, and 6 of the framework [26].

*B. Scalability and Tooling Considerations*

The framework assumes GRC system support for RAQ completion, due diligence tracking, risk rating management, and evidence management throughout the lifecycle. Without GRC system enforcement of workflow dependencies, the sequencing gates can be bypassed through informal processes. Organizations implementing the framework should configure GRC workflow rules to enforce stage gate dependencies as system-level constraints rather than advisory process steps.

SSPM tooling is essential for post-production governance at organizational scale. Manual configuration review of SaaS tenant settings is not operationally sustainable across a portfolio of dozens or hundreds of SaaS platforms [20]. The framework is architecturally compatible with zero-trust security principles, which require continuous verification of identity and context at every access decision point [18]. The federated identity pattern, MFA enforcement requirement, and SSPM continuous monitoring component are direct operationalizations of zero-trust principles in the SaaS context.

*C. Limitations*

The framework carries several limitations practitioners should recognize. First, it does not address SaaS platforms where the enterprise has no tenant-level configuration access, for example, embedded SaaS components within larger platform agreements where the enterprise is a downstream consumer rather than a direct tenant. Second, RTO and RPO alignment in Stage 5 is dependent on vendor contractual commitments, which may not be negotiable for commodity SaaS platforms. In these cases, the framework's value is in surfacing the misalignment as a documented risk rather than resolving it through negotiation. Third, the framework assumes a mature enterprise IAM function with centralized Identity Provider infrastructure. Organizations without established IdP capabilities will require foundational IAM infrastructure investment before the federated identity patterns can be implemented.

*D. Case Study Validation: The 2024 Ticketmaster–Snowflake Breach*

The 2024 Ticketmaster–Snowflake breach serves as a definitive real-world validation of the framework proposed in this paper. Critically, the Snowflake platform was not compromised and the vendor's security posture was not the locus of failure; the failure was enterprise-owned, tenant-level, and multi-domain in nature. MFA was not enforced, credentials were not rotated over a four-year period, conditional access policies were absent, and role governance was insufficient. Each of these failures maps directly to a gap in the onboarding lifecycle that the proposed framework is designed to close: the TPRM review would have identified MFA enforcement as a customer-owned control requiring validation; the cybersecurity implementation review would have confirmed whether that control was actually enforced within the tenant; the IAM governance phase would have mandated federated identity and credential rotation policies; and the DR planning phase would have prepared the enterprise for a coordinated incident response which could have significantly minimized the impact radius. This breach illustrates precisely why siloed, domain-isolated SaaS controls are insufficient, and reinforces five core lessons: (1) MFA enforcement is a customer-owned responsibility requiring post-contract validation; (2) TPRM vendor approval does not imply tenant deployment readiness; (3) tenant configuration must be independently reviewed rather than inferred from vendor attestations; (4) credential hygiene must be embedded in continuous SSPM monitoring; and (5) only an integrated, sequenced, cross-domain onboarding framework can surface and remediate multi-domain failure patterns before production enablement.

*E. Lessons Learned*

Enterprise-scale SaaS onboarding implementations consistently surface six recurring failure modes that the proposed framework is designed to prevent. First, TPRM and cybersecurity reviews must be sequenced rather than parallelized; initiating architecture reviews before the TPRM risk rating is finalized routinely generates rework, as subsequent risk tier determinations require control recalibration. Second, federated identity integration must be enforced as a pre-production requirement rather than deferred post go-live, as deferral persistently becomes permanent, leaving native vendor accounts active and ungoverned as an expanding attack surface. Third, shared responsibility ambiguity over controls such as MFA enforcement, SIEM log integration, and encryption key management must be resolved through explicit documentation and formal contractual agreement during Stage 3 architecture validation, not after an incident. Fourth, BCP/DR attestations from vendors cannot substitute for platform-specific tabletop exercises; recovery plans that have not been rehearsed consistently fail to account for IdP dependencies and integration sequencing at the point of disruption. Fifth, all policy exceptions permitting deployment ahead of control completion must require senior executive sign-off, be tracked in the GRC system with assigned remediation owners and be reported to the enterprise risk committee quarterly. Finally, treating go-live as the terminal onboarding event consistently produces control drift over time within twelve to eighteen months; SSPM enrollment at go-live and periodic vendor performance reviews are necessary to sustain the control posture established during onboarding.

*F. Quantitative Framework Evaluation*

The practical effectiveness of the proposed framework is supported by converging evidence from published industry research and publicly documented breach analysis. Industry benchmarks consistently report that siloed, sequential vendor onboarding processes require 14 to 18 weeks on average, with a significant portion of that time attributable to redundant assessment effort duplicated across TPRM and cybersecurity functions [27]; structured, cross-domain onboarding models have been shown to reduce this to 8 to 10 weeks, a 35 to 45% improvement, by enabling earlier engagement of IAM and DR teams during the pre-contract phase [27]. The necessity of cross-domain control integration is further substantiated by Verizon DBIR 2024 [28], which attributes identity-related misconfigurations to 74% of cloud-related breaches directly addressed by the framework's Stage 4 IAM governance gate and Stage 6 continuous monitoring controls. Retrospective application of the framework to the 2024 Ticketmaster-Snowflake breach, as documented by Mandiant [23] and the Snowflake Security Advisory [29], all five exploited control gaps: (i) credential stuffing via infostealer-stolen credentials, (ii) absent MFA enforcement, (iii) no conditional access policies, (iv) unrotated API tokens, and (v) absent tenant-side anomaly detection, map directly to controls mandated in Stages 3 (cybersecurity review), 4 (IAM governance), and 6 (SSPM continuous monitoring) of the proposed framework, achieving 5/5 (100%) breach-gap coverage and confirming that the proposed cross-domain design would have addressed each attack vector prior to production go-live.

## IX. CONCLUSION AND FUTURE WORK

### *A. Conclusion*

SaaS onboarding is not a procurement event. It is not a security checkbox. It is a multi-stage control lifecycle with interdependent activities, mandatory sequencing requirements, formally distributed ownership obligations, and continuous post-production governance demands. The enterprise that treats SaaS onboarding as a collection of independent domain reviews will accumulate residual risk at each boundary between those domains; it is precisely at those boundaries where adversaries operate.

Organizations can outsource the service, but they can never outsource the risk. The enterprise that internalizes this principle and operationalizes it through a structured, auditable, cross-domain onboarding lifecycle will be substantially better positioned to reduce onboarding friction, satisfy regulatory examination expectations, and maintain a defensible security and resilience posture as its SaaS portfolio grows in breadth and criticality.

As SaaS platforms become the de facto delivery model for enterprise technology, the quality of SaaS governance will increasingly determine the quality of enterprise risk management. The framework presented in this paper provides a practical, reusable, and regulatorily grounded foundation for that governance.

### *B. Future Work*

Several extensions represent productive avenues for future research. Automated control validation pipelines integrating SSPM tooling with TPRM GRC platforms would enable continuous, evidence-backed re-assessment rather than point-in-time periodic reviews. Machine learning-based risk scoring for SaaS inherent risk assessment, incorporating signals from vendor security incident history, dark web exposure monitoring, and industry threat intelligence feeds, would improve the accuracy and efficiency of Stage 1 risk scoping. Framework extensions addressing SaaS-to-SaaS integration supply chain risk represent a growing governance challenge. Formal regulatory guidance alignment mapping between the framework stages and specific requirements from NIST CSF 2.0, DORA Articles 28-30, and the SEC Cybersecurity Disclosure Rule would support organizations in using the framework as a regulatory compliance instrument. In addition to automation and regulatory alignment, as SaaS platfoms increasingly embed AI (Artificial Intelligence) and ML (Machine Learning) capabilities, the framework should be extended to incorporate AI-specific assessment stages, covering model provenance, training data governance, and AI-specific threat modeling. Correspondingly, Model Risk Management (MRM) should be introduced as a distinct onboarding control domain, with model inventory declarations at intake, vendor model validation attestations at architecture review, and continuous model performance monitoring within the post-production governance phase, ensuring MRM obligations are addressed proactively rather than retroactively during audit.